# Gene selection from microarray expression data A Multi-objective PSO with adaptive K-nearest neighborhood

Yasamin Kowsari
Department of Computer Engineering
Azad University of Mashhad
Mashhad, Iran
Kowsariyasamin@gmail.com

Sanaz Nakhodchi
Department of Computer Engineering
University of Guelph
Guelph, Ontario, Canada
Email: nakhodcs@uoguelph.ca

Davoud Gholamiangonabadi
Department of Computer Engineering
Western University
London, Ontario, Canada
dgholamian@gmail.com

*Abstract*— Cancer detection is one of the key research topics in the medical field. Accurate detection of different cancer types is valuable in providing better treatment facilities and risk minimization for patients. This paper deals with the classification problem of human cancer diseases by using gene expression data. It is presented a new methodology to analyze microarray datasets and efficiently classify cancer diseases. The new method first employs Signal to Noise Ratio (SNR) to find a list of a small subset of non-redundant genes. Then, after normalization, it is used Multi-Objective Particle Swarm Optimization (MOPSO) for feature selection and employed Adaptive K-Nearest Neighborhood (KNN) for cancer disease classification. This method improves the classification accuracy of cancer classification by reducing the number of features. The proposed methodology is evaluated by classifying cancer diseases in five cancer datasets. The results are compared with the most recent approaches, which increases the classification accuracy in each dataset.

## I. INTRODUCTION

Gene expression data are increasingly being used in medicine and machine learning [1][2]. With the help of different improvements in the field of machine learning, different applications in the medical field are employed, such as brain-computer interface (BCI) [3], cancer detection by using different types of clinical images [4], and drug discovery [5], robotics[6] and lot of different areas. An important research field in bioinformatics and cancer detection is the classification of gene expression data [7]. Feature selection is a multi-objective optimization problem. It has two conflicting objectives: maximizing the classification accuracy and minimizing the number of the selected features among the vast number of genes [8]. A hybrid feature selection algorithm is proposed in [9], and combines the mutual information maximization (MIM) and the adaptive genetic algorithm (AGA). Experimental results show that the MIMAGA-Selection method significantly reduces the dimension of gene expression data and removes the redundancies for classification. A new methodology combines both Information Gain (IG) and Standard Genetic Algorithm (SGA) introduced in [10]. It first uses Information Gain for feature selection, then uses Genetic Algorithm (GA) for feature reduction, and finally uses Genetic Programming (GP) for cancer type classification. The suggested system is evaluated by classifying cancer diseases in seven cancer datasets, and the results are compared with the latest approaches. A rough-based hybrid binary PSO algorithm was proposed in [11], which uses a heuristic-based fast processing strategy to reduce features by eliminating redundant features and then discretizing subsequently into a binary table, known as a distinction table. This distinction table is used to optimize the cost functions to generate a reduction in rough set theory. Another method refines the feature (gene) space from a very coarse level to a fine-grained one at each recursive step of the algorithm without degrading the accuracy. In addition, it integrated various filter-based ranking methods with the recursive PSO approach [12]. Ant Colony Optimization (ACO) and Ant Lion Optimization (ALO) algorithm are proposed and employed in the muted selection process. The ant lions to hunt process is proposed for Leukemia prediction using microarray gene data[13]. For identifying the irrelevant and the redundant features, an adopted FCBF (First Correlation-based feature selection) is used in [14]. in the next step, by considering SVM and main algorithms PSO and recursive FA (Firefly algorithm) optimization process finished, the adopted method is known as PRFA-SVM. The two significant objects, such as achieving higher accuracy and a small number of features, make the research challenging. A Recursive Memetic Algorithm (RMA) algorithm for selecting genes is introduced in [15]. This method has improved the Memetic Algorithm (MA) and has better results than MA and the Genetic Algorithm (GA). A hybrid technique for gene selection, called ensemble multi-population adaptive genetic algorithm (EMPAGA) proposed in [16], can overlook the irrelevant genes and classify cancer accurately. A support vector machine and naive Bayes are applied to select the essential genes and classify them in an actual class. A novel PSO-based multi-objective feature selection method was proposed in [17]. This methodology consists of three steps. At first, original features are shown as a graph representation model. In the next step, feature centralities for all nodes in the graph are calculated, and finally, in the third phase, an improved PSO-based search process is utilized for the final feature selection. A two-stage gene selection by applying extreme gradient boosting (XGBoost) is proposed in [18]. Also, a multi-objective optimization genetic algorithm (XGBoost-MOGA) for cancer classification in microarray datasets is introduced.

## II. PROPOSED METHOD

In this method, we have three steps fig. 1 shows a summary of their operations. In the first step, we have a preprocess function, then, in the second step, we explain MOPSO, and in the third step, we discuss the adaptive KNN. Selected genes were set on a minimum of ten, and then an adaptive algorithm to maximize the classification accuracy was applied.

### A. Preprocessing

In this research, The dataset is an extensive matrix of microarray data in which the first column of data shows types of disease, and the other columns are the genes of the data matrix, according to the decreasing order of obtained SNR and normalization function.

SNR: The SNR explains the ratio between the relative mean and the sum of the Standard Deviation of two classes of different samples. A low SNR indicates that the feature does not have much different information in different classes. In contrast, high SNR indicates that the feature values are over an extensive range of information, and it is expected that the values are different in all classes. The SNR for each feature or gene is calculated, and it is seen that some features have very low SNR that may be considered to be insignificant to the class labels and are filtered out before further experiments. It is also noticed that for all the different data sets after features are sorted according to descending order of SNR, the top few features perform well. We have chosen the top 100 features as it provides a tradeoff between the time requirement and the performance of the algorithms [2],[10],[19],[20]. The equation of SNR is in (1).

$$|SNR| = \left| \frac{mean(class1) - mean(class2)}{SD(class1) + SD(class2)} \right| \quad (1)$$

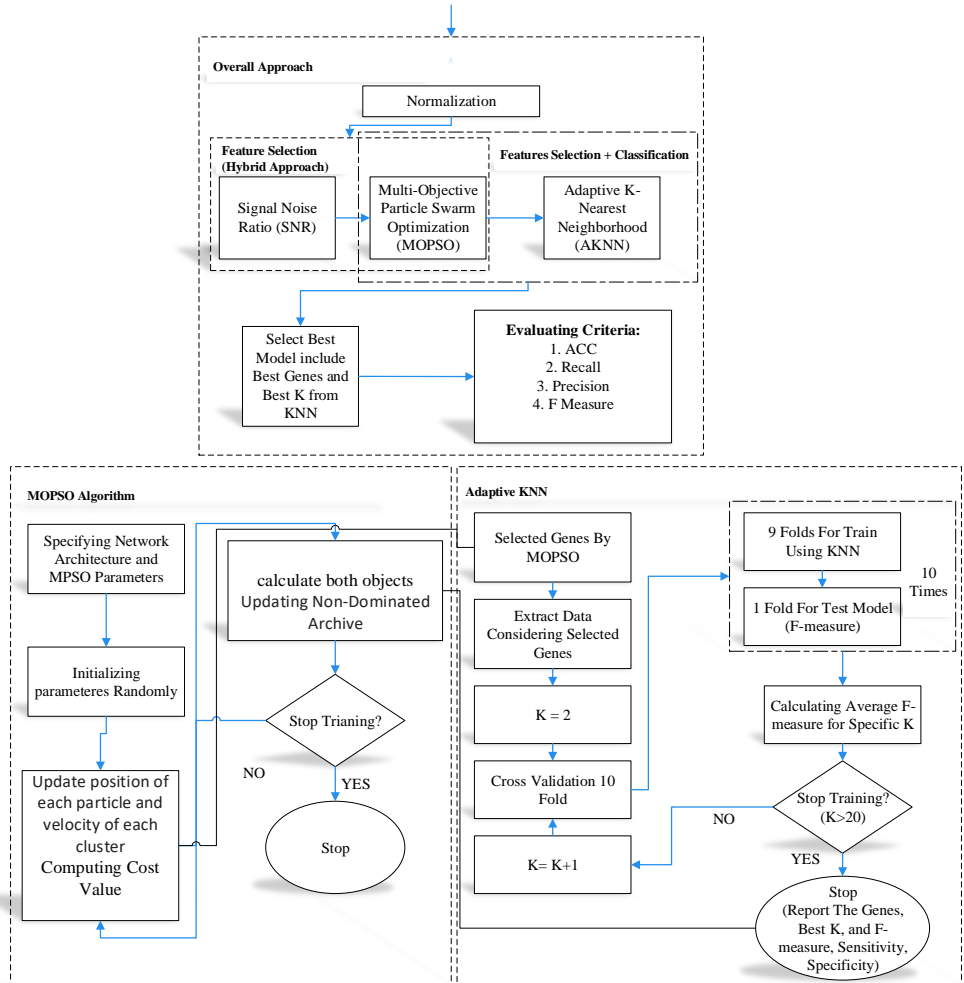

Fig. 1 An overview of the implementation process

*Normalization:* The data matrix is normalized to set each gene expression value from 0 to 1. For normalization, minimum and maximum values of each gene (column) are calculated first. Then normalization is done by Equation (2). where gij denotes gene expression value of ith sample of the jth gene and gj denotes the gene expression values of jth gene across all the samples. Undoubtedly, considering more genes as input creates more options for selecting a suitable set of features (genes) that improves classification accuracy. For this article, we consider only 100 genes as input which is an appealing choice for classification[2][17][21].

$$Normaliz(g_{ij}) = \frac{g_{ij} - minimum(g_j)}{maximum(g_j) - minimum(g_j)} \quad (2)$$

### B. MOPSO

The particle reformed is defined in fig. 2 (a), which shows a particle with 2ŋ cells. The first ŋ cells by values between (0,1) and other ŋ cells represent ŋ cluster centers or genes. If padding cell(i) >threshold, gene ith is selected for fitness computation.

The second part of the candidate solution, fig.2 (b), shows ŋ cluster centers (CC). Each cluster consists of one gene representing the cluster center. Vij represents a velocity of ith candidate solution and jth cluster. A gene contains S number of samples. Xijk corresponds to a position of ith particle jth cluster and kth sample. Predicted class labels are compared with the original class labels.

In this proposed method, multi-objective particle swarm optimization (MOPSO) has been introduced to maximize the value of sensitivity and specificity to select significant features. The number of essential measures is shown as false positives (fp), true negatives (tn), false negatives (fn), and true positives (tp). After using these four terms, sensitivity (3) and specificity (4) are determined.

$$Sensitivity = \frac{(tp)}{(tp + fn)} \quad (3)$$

$$Specificity = \frac{(tn)}{(tn + fp)} \quad (4)$$

The initializing of the population is by randomly chosen features from the data matrix, and the corresponding fitness values are calculated by adaptive KNN using k-fold cross-validation. The archive A is initialized by solutions in the non-dominated model from the population after applying the non-domination function to the initial population. Velocity and position are updated using Equations (5).

$$V_{nd} = wv_{nd} + r1(P_{nd} - X_{nd}) + r2(G_{nd} - X_{nd}).$$
$$X_{nd} = X_{nd} + V_{nd}.$$
$$X_{nd} = CellBoundary(x_{nd}). \quad (5)$$

CellBoundary checks that the value of each cell from the first part after the update is in the range of 0 to 1, and if it is out of range, we replace it with the value of the nearest boundary. After we update the position and velocity for each particle, DimensionBoundary is a function that corrects genes within the range if it exceeds the allowable range. When this method is applied to the data matrix, a set of non-dominated candidate solutions is obtained after updating the cluster centers located in the second part of the particle N based on the cluster center selection rules. After updating all the particles, then merge the archive and the new particle population, and finally build the archive based on the concept of overcoming. Since the size of the archive is limited (it means that only 30 unsuccessful solutions can be placed in the archive), if the number of solutions in the archive is more than the size of the archive, then remove a number of solutions from the archive and increase the number of solutions in the archive to the number of allowed solutions in the archive. We use the crowding criterion to determine which solution is the target. First, we calculate this criterion for each solution in the archive, and then we keep the solutions for which the value of this criterion is larger in the archive. Next, the F-score measure is computed for selecting one from these non-dominated candidate solutions. The F-score measure combines precision and recall through the harmonic mean of precision and recall shown in Equations (6) and (7).

$$F = \frac{2 * Precision * Recall}{Precision + Recall} \quad (6)$$

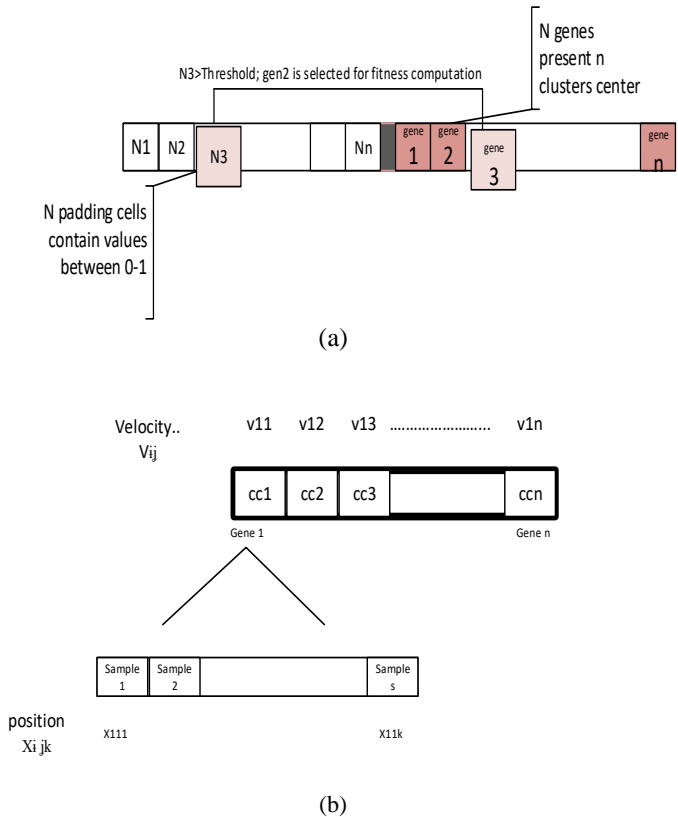

Fig. 2 Demonstration of particle encoding

$$precision = \frac{tp}{tp + fp} \quad (7)$$

*C. Adaptive KNN*

For classifying each sample, the Adaptive KNN method was used. For this goal, k fold cross-validation method has been applied. Folded ten times to select training and test data [1]. In cross-validation, the data of one of the fields are used as a set of evaluators (test), and the other (k-1) folds are used as training data. In each iteration of the experiment, the K value of the KNN algorithm changes from 3 to 20. After this step, we calculate the accuracy of test data and training data for cross-validation by using Equations (8) and (9), and (10). Cross-validation training accuracy (CV training accuracy) and cross-validation test accuracy (CV test accuracy) are calculated. After this step, Tr and Ts are defined as the training accuracy and test accuracy for each fold. For each K, K numbers of single estimations (SE) are calculated. From that K number of single estimations, an optimal value is selected. In our case, the optimal value is chosen as: max {SE(K)}. This optimal value of single estimation is named cross-validation accuracy (CV accuracy), and the corresponding K value for the KNN classification method is selected as Kopt or optimal k. Selection of Kopt and CV accuracy is computed for each particle for all the iterations of the MOPSO–adaptive KNN algorithm.

$$CV\ Training\ accuracy = \frac{1}{k}\sum_{m=1}^{l} Tr_m \quad (8)$$

$$CV\ Test\ accuracy = \frac{1}{k}\sum_{m=1}^{l} Ts_m \quad (9)$$

$$SE(k) = \frac{1}{2}\left(\frac{1}{k}\sum_{m=1}^{l} Tr_m + \frac{1}{k}\sum_{m=1}^{l} Ts_m\right) \quad (10)$$

III. EXPERIMENTAL RESULTS

*A. Datasets*

In this research we choose two class datasets in the field and the features of every dataset is described in below.

- DLBCL: B-cell lymphomas (DLBCL) and follicular lymphomas (FL) are two B-cell lineage malignancies that have variety clinical presentations, This data can be acquired from the website: www.biolab.si/supp/bi-cancer/projections/info/DLBCL.htm/. Total 7,070 genes are there in the data set. The number of samples is 58 and type FL is 19[2],[15],[16].
- Child_ALL: The childhood ALL data set (GSE412) includes gene expression information in 110 childhood acute lymphoblastic leukemia samples. The data set describes childhood acute lymphoblastic leukemia cells based on changes in gene expression before and after treatment, regardless of the type of treatment used. This data set was collected from the same website: www.biolab.si/supp/bi-cancer/projections/info/ GSE412.htm/. The data set has 50 examples of type before therapy and 60 examples of type after therapy. The number of genes are 8,280.
- Prostate: Gene expression measurements for samples of prostate tumors and adjacent prostate tissue not containing tumor were used to build this classification model. It contains 50 normal tissues and 52 prostate tumor samples. The expression matrix consists of 12,533 genes and 102 samples. This data is available at the following website: www.biolab.si/supp/bi cancer/projections/info/prostata.htm/[2],[15],[16].
- Chornic myeloid leukemia: The majority of patients with chronic myeloid leukemia (CML) in chronic phase. However, a subgroup of patients doesn't respond to standard treatment with imatinib. Clinically, it would be advantageous to identify such patients in advance, since they may benefit from more aggressive therapy. There are 12 nonresponders and 16 responders to imatinib treatment. Therefore, the total number of samples is 28 and this data set contains 12,625 genes. This data is available at the following website: www.biolab.si/supp/ [2],[15],[16].
- Colon cancer: the total number of samples is 62 and this data set contains 2000 genes.

*B. Evaluations Measures*

In order to assess the solution quality of our method, four evaluation measures including precision, recall, F-score, and accuracy were showed (3), (4), (6), (11).

$$accuracy = \frac{(tp + tn)}{(tp + tn + fp + fn)} \quad (11)$$

*C. Discussion*

In this section, we first examine the said criteria in Table I in five different parts, and each one is related to one dataset. We also review the results of the implementation of the proposed method with previous articles for all five datasets. It is noteworthy that the results presented from previous articles are either average performance or expressed by the best and average and worst implementation accuracy, the average for the proposed method indicates the amount of accuracy after 10 times the implementation. The results of the implementation in Table II show that the proposed method has high accuracy in all five datasets, and the criteria expressed individually show the high capability of this method in classifying genomic data expression microarrays.

Table I   The percentage of average performance of classifiers on 5 gene datasets

| Accuracy | sensitivity | specificity | F-score | method |
|---|---|---|---|---|
| colspan=5 | DLBCL | | | |
| 90 | 90 | 90 | 90 | GA adaptive SVM without preprocess |
| 99 | 100 | 98 | 99 | This method |
| colspan=5 | Child ALL | | | |
| 94 | 92 | 97 | 94 | Ant colony[13] |
| 94 | 90 | 1 | 95 | This method |
| colspan=5 | Prostate | | | |
| 93 | 92 | 93 | 92 | GA adaptive SVM without preprocess[16] |
| 100 | 100 | 100 | 100 | This method |
| colspan=5 | Chornic myeloid leukemia | | | |
| 93 | 90 | 90 | 89 | GA adaptive SVM without preprocess[16] |
| 100 | 100 | 100 | 100 | This method |
| colspan=5 | Colon cancer | | | |
| 93 | 94 | 93 | 90 | GA adaptive SVM without preprocess[16] |
| 98 | 1 | 98 | 98 | This method |

TABLE II   The percentage of average accuracy of classifiers on 5 gene datasets

| Method | datasets | | | | |
|---|---|---|---|---|---|
| | Chornic myeloid leukemia | DLBCL | Child_ALL | Prostate | Colon cancer |
| Information Gain (IG) and Standard Genetic[10] | 91.17 | 88.3 | ---- | 100 | 85.48 |
| Ant colony [13] | --- | ---- | 93.94 | ---- | 91.93 |
| Multi objective pso[17] | 90.16 | ---- | ---- | 82.81 | |
| PRFA svm [14] | 100 | ---- | ---- | 100 | 96.2 |
| Adaptive GA [16] | Best:98.99 Avg: 98.53 Min:98.26 | Best:100 Avg: 99.23 Min: 96.83 | ---- | Best: 98.62 Avg: 97.23 Min:90.52 | Best: 98.63 Avg: 97.23 Min:94.06 |
| Recursive memetic [15] | 94.1 | 97.2 | ---- | 92.8 | 100 |
| Multi objective PSO[2] | ------ | 80.35 | 81.45 | 92.35 | ---- |
| XGBoost-MOGA[18] | …….. | 100 | …… | 98.00 | 90.24 |
| SNR and Kmeans and colony optimization[19] | ……… | 99.42 | ….. | 96.47 | ……. |
| **Perposed method** | Best:100 Avg: 100 Min:100 | Best:100 Avg:99.47 Min:98.32 | Best:96.65 Avg: 94.62 Min:92.91 | Best:100 Avg: 100 Min:100 | Best:100 Avg: 98.81 Min:98.32 |

## IV. Conclusion

The importance of diagnosing the disease from genomic data is undeniable. Genomic microarrays play a vital role in diagnosing some diseases, and various datasets have been collected to diagnose different types of disorders and cancerous tumors. Each of these datasets expresses different information about genes and in different subclasses. In this study, five datasets of two-class datasets of genomic expression microarrays were investigated. The primary purpose was to increase the identification of the appropriate subgroup. The main challenge is the large size of the microarrays, which makes it challenging to identify the suitable subset of genes to diagnose the disease subset. In this research, after examining the previous methods, by focusing on multi-objective algorithms, an attempt has been made to create more solutions. A multi-objective particle swarm optimization algorithm suggests a more desirable solution in presenting and selecting the most effective available genes. Also, by using the K nearest neighborhood algorithm and using it in classifying the appropriate subgroup of the disease adaptively, the accuracy of classification has increased. The proposed method was tested with the Prostate, DLBCL, Chronic Myeloid Leukemia, colon cancer, and All Child datasets, and the results were presented with other suggested ideas for better accuracy.